\documentclass[aps,prd,showpacs,nofootinbib,floatfix,twocolumn]{revtex4}
\usepackage{amsmath,latexsym,graphicx}
\begin{document}
\title{On the falloff of radiated energy in black hole spacetimes}

\author{Lior M.\ Burko$^{1,2}$ and Scott A.\ Hughes$^{3}$}

\affiliation{$^1$Department of Physics, University of Alabama in
Huntsville, Huntsville, Alabama 35899}
\affiliation{$^2$Center for Space Plasma and Aeronomic Research,
University of Alabama in Huntsville, Huntsville, Alabama 35899}
\affiliation{$^3$Department of Physics and MIT Kavli Institute, MIT,
77 Massachusetts Ave., Cambridge, MA 02139}
\begin{abstract}
The goal of much research in relativity is to understand gravitational
waves generated by a strong-field dynamical spacetime.  Quantities of
particular interest for many calculations are the Weyl scalar
$\psi_4$, which is simply related to the flux of gravitational waves
far from the source, and the flux of energy carried to distant
observers, $\dot E$.  Conservation laws guarantee that, in
asympotically flat spacetimes, $\psi_4 \propto 1/r$ and $\dot E
\propto 1/r^2$ as $r \to \infty$.  Most calculations extract these
quantities at some finite extraction radius.  An understanding of
finite radius corrections to $\psi_4$ and $\dot E$ allows us to more
accurately infer their asymptotic values from a computation.  In this
paper, we show that, if the final state of the system is a black hole,
then the leading correction to $\psi_4$ is ${\cal O}(1/r^3)$, and that
to the energy flux is ${\cal O}(1/r^4)$ --- not ${\cal O}(1/r^2)$ and
${\cal O}(1/r^3)$ as one might naively guess.  Our argument only
relies on the behavior of the curvature scalars for black hole
spacetimes.  Using black hole perturbation theory, we calculate the
corrections to the leading falloff, showing that it is quite easy to
correct for finite extraction radius effects.
\end{abstract}
\maketitle

\section{Introduction}
\label{sec:intro}

Extracting radiation from the output of numerical calculations, as
well as fluxes of quantities such as energy carried by radiation, is
important for many problems in general relativity.  Newman \& Unti
{\cite{nu62}} provide an outstanding foundation for understanding
analytically the asymptotic behavior of curvature tensors which
determine how gravitational radiation behaves as it propagates far
from a radiating source.  Perturbation theory also provides an
excellent set of tools to help us understand the asymptotic behavior
of radiation and fluxes.

Many results on the distant behavior of radiation fields describe how
quantities behave in the limit $r \to \infty$.  With the exception of
characteristic methods (see, for example, {\cite{winicour}}), most
numerical calculations extract radiation at some large but finite
radius $r$.  Understanding the subleading corrections to the
asymptotic behavior of radiative quantities could greatly improve our
ability to extract asymptotic fluxes and fields from numerical codes.

Previous work {\cite{skh07}} found empirically that the form
\begin{equation}
\dot E(r) = \dot E_\infty\left(1 + \frac{e_2}{r^2}\right)
\label{eq:edot_subleading}
\end{equation}
does an outstanding job describing subleading corrections to the
gravitational-wave energy flux.  In this paper, we examine this
behavior more carefully.  In Sec.\ {\ref{sec:analytic}}, following the
formalism developed in Ref.\ {\cite{nu62}}, we prove that this form is
to be generically expected, and follows from the fact that at finite
large radius $r$, the Weyl curvature scalar describing distant
radiation takes the form $\psi_4(r) = \psi_4^\infty\left(1 +
b_2/r^2\right)$.  In Sec.\ {\ref{sec:perturb}}, we use black hole
perturbation theory to calculate the coefficients $b_2$ and $e_2$.  We
conclude Sec.\ {\ref{sec:discuss}} by discussing possible applications
of this result.

\section{Tools and formalism for understanding radiation falloff}
\label{sec:analytic}

\subsection{Definitions}

We begin by defining the quantities which we will need for our
analysis.  Much of this discussion is adapted from Ref.\
{\cite{nu62}}.  We present these general definitions in some detail
before specializing to the much simpler black hole case.

Consider a vacuum, asymptotically flat spacetime.  Introduce a family
of null hypersurfaces, each characterized by a constant parameter $u$.
We take $u = x^0$ as one of the coordinates we will use to describe
our geometry.  Define
\begin{equation}
l_\alpha = \partial_\alpha u\;.
\label{eq:ldef}
\end{equation}
Since these surfaces are null, the vector $l^\alpha$ is tangent to
null geodesics.  This vector will be the first leg of a tetrad which
we will use to characterize our geometry.  Define $r$ as the affine
parameter along these geodesics; this will denote another of our
coordinates.  The remaining coordinates $x^k$ ($k \in 3,4$) then label
the different null geodesics in each constant $u$ hypersurface; they
can be taken to be angles.

We define a second null vector $n^\alpha$ by requiring
\begin{equation}
n^\alpha l_\alpha = 1\;.
\label{eq:ndef}
\end{equation}
To complete our tetrad, we next define a pair of unit spacelike
vectors $\zeta^\alpha$ and $\rho^\alpha$ that are orthogonal to
$l^\alpha$, $n^\alpha$, and each other.  We then put
\begin{eqnarray}
m^\alpha = (\zeta^\alpha - i \rho^\alpha)/\sqrt{2}\;,
\label{eq:mdef}
\\
{\bar m}^\alpha = (\zeta^\alpha + i \rho^\alpha)/\sqrt{2}\;.
\label{eq:mbardef}
\end{eqnarray}

We now use this tetrad to characterize the curvature of our spacetime.
Let $C_{\alpha\mu\beta\nu}$ be the Weyl (vacuum) curvature tensor of
the spacetime.  Define the following 5 complex Weyl projections:
\begin{eqnarray}
\psi_0 &=& -C_{\alpha\mu\beta\nu}l^\alpha m^\mu l^\beta m^\nu\;,
\label{eq:psi0}
\\
\psi_1 &=& -C_{\alpha\mu\beta\nu}l^\alpha n^\mu l^\beta m^\nu\;,
\label{eq:psi1}
\\
\psi_2 &=& -C_{\alpha\mu\beta\nu}l^\alpha m^\mu {\bar m}^\beta
n^\nu\;,
\label{eq:psi2}
\\
\psi_3 &=& -C_{\alpha\mu\beta\nu}l^\alpha n^\mu {\bar m}^\beta
n^\nu\;,
\label{eq:psi3}
\\
\psi_4 &=& -C_{\alpha\mu\beta\nu}n^\alpha {\bar m}^\mu n^\beta {\bar
m}^\nu\;.
\label{eq:psi4}
\end{eqnarray}

Reference {\cite{nu62}} shows that as we approach the asymptotically
flat ($r \to \infty$) regime, these curvature components vary as
follows:
\begin{eqnarray}
\psi_0 &=& \frac{A_0}{r^5} + {\cal O}(1/r^6)\;,
\label{eq:psi0_large_r}
\\
\psi_1 &=& \frac{A_1}{r^4} + \frac{\left(4\alpha_{\rm RSC} A_0 -
\bar\xi^k\partial_k A_0\right)}{r^5} + {\cal O}(1/r^6)\;,
\label{eq:psi1_large_r}
\\
\psi_2 &=& \frac{A_2}{r^3} + \frac{\left(2\alpha_{\rm RSC} A_1 -
\bar\xi^k\partial_k A_1\right)}{r^4} + {\cal O}(1/r^5)\;,
\label{eq:psi2_large_r}
\\
\psi_3 &=& \frac{A_3}{r^2} - \frac{\bar\xi^k\partial_k A_2}{r^3} +
{\cal O}(1/r^4)\;,
\label{eq:psi3_large_r}
\\
\psi_4 &=& \frac{A_4}{r} - \frac{\left(2\alpha_{\rm RSC} A_3 +
\bar\xi^k\partial_k A_3\right)}{r^2} + {\cal O}(1/r^3)\;.
\label{eq:psi4_large_r}
\end{eqnarray}
In Eqs.\ (\ref{eq:psi1_large_r}) -- (\ref{eq:psi4_large_r}), the index
$k \in [3,4]$, the complex function $\xi^k$ describes the angular
components of the tetrad element $m^\alpha$, and the functions
$\alpha_{\rm RSC}$ and $\gamma_{\rm RSC}$ are ``Ricci spin
coefficients,'' constructed by certain combinations and projections of
the tetrad's covariant derivatives.  For more details and discussion
of these functions, see Refs.\ {\cite{nu62,np62}}.  For our purposes,
the most important fact to take from Eqs.\ (\ref{eq:psi0_large_r}) --
(\ref{eq:psi4_large_r}) is that the leading falloff of $\psi_4$ is at
${\cal O}(1/r)$.  The subleading correction at ${\cal O}(1/r^2)$ is
set by a coefficient that scales with $A_3$, which controls the
behavior of the curvature scalar $\psi_3$.

\subsection{Perturbed black holes}

We now specialize to black holes.  We use the Kinnersley tetrad
{\cite{k69}}, which in Boyer-Lindquist coordinates is given by
\begin{eqnarray}
l^\alpha &\doteq& \frac{1}{\Delta}\left(r^2 + a^2, 1, 0, a\right)\;,
\\
n^\alpha &\doteq& \frac{1}{2\Sigma}\left(r^2 + a^2, -\Delta, 0, a\right)\;,
\\
m^\alpha &\doteq& \frac{1}{\sqrt{2}(r + ia\cos\theta)}
\left(ia\sin\theta, 0, 1, i\csc\theta\right)\;.
\end{eqnarray}
For an unperturbed black hole spacetime, $\psi_2 = -M/(r -
ia\cos\theta)^3$, and $\psi_n = 0$ for $n \ne 2$.  Far from a {\it
perturbed} black hole, $\psi_4$ is also non-zero, describing the
spacetime's outgoing gravitational waves:
\begin{eqnarray}
\psi_4(r \to \infty) &=& \frac{1}{2}\left(\ddot h_+ - i\ddot
h_\times\right)
\nonumber\\
&=& \frac{1}{2r}\left(\ddot H_+ - i\ddot H_\times\right)\;.
\label{eq:psi4_waves}
\end{eqnarray}
The Weyl scalar $\psi_0$ is also generically non-zero for a perturbed
black hole, but we will not need its value in our analysis.  Crucially
for our argument, we can always put $\Psi_3 = 0$ for our perturbed
black hole {\cite{bb02}}.

Comparing with Eqs.\ (\ref{eq:psi1_large_r}) --
(\ref{eq:psi4_large_r}), we read off
\begin{eqnarray}
A_3 &=& 0\;,
\\
A_2 &=& -M\;,
\\
A_4 &=& \frac{1}{2}\left(\ddot H_+ - i\ddot H_\times\right)\;.
\end{eqnarray}
Combining these results with Eq.\ (\ref{eq:psi4_large_r}), we see that
corrections to $\psi_4$ come in at ${\cal O}(1/r^3)$, so that
\begin{equation}
\psi_4 = \frac{A_4}{r}\left(1 + \frac{b_2}{r^2}\right)\;,
\label{eq:psi4_subleading}
\end{equation}
where $b_2$ is a complex constant related to the (currently unknown)
coefficient of this subleading falloff.

\subsection{Energy flux}

We now relate the curvature scalar $\psi_4$ to the asymptotic flux of
radiation from the source.  The energy flux in gravitational waves is
given by
\begin{equation}
\dot E = \frac{1}{16\pi}\int r^2d\Omega\left[(\dot h_+)^2 + (\dot
h_\times)^2\right]\;.
\label{eq:edot_h}
\end{equation}
Using Eq.\ (\ref{eq:psi4_waves}), we can relate this to $\psi_4$ in
the limit $r \to \infty$:
\begin{equation}
\dot E^\infty = \frac{1}{4\pi}\lim_{r\to\infty}\int
r^2d\Omega\,\biggl|\int dt\,\psi_4 \biggr|^2\;.
\label{eq:edot_psi4}
\end{equation}
Using Eq.\ (\ref{eq:psi4_subleading}), let us now see what this
implies about the behavior of $\dot E$ when radiation is extracted at
some finite radius $R$.  Let us first introduce a modal expansion,
writing
\begin{eqnarray}
\psi_4 &=& \sum_\omega \psi_4^\omega e^{-i\omega t}
\nonumber\\
&=& \sum_\omega \frac{A_4^\omega}{r}\left(1 +
\frac{b_2^\omega}{r^2}\right) e^{-i\omega t}
\label{eq:psi4_timedecomp}
\end{eqnarray}
For simplicity, we have taken the radiation to have a discrete
frequency spectrum.  The calculation can easily be extended to
encompass a continuous spectrum.  Combining Eqs.\ (\ref{eq:edot_psi4})
and (\ref{eq:psi4_timedecomp}), we find
\begin{eqnarray}
\dot E(r) &=& \frac{1}{4\pi}\sum_\omega \omega^{-2} \int
r^2\,d\Omega\,|\psi_4^\omega|^2
\label{eq:edot_psi}
\\
&=& \sum_\omega \dot E^\omega_\infty \left(1 +
\frac{e_2^\omega}{r^2}\right)\;,
\label{eq:edot_correction}
\end{eqnarray}
where
\begin{equation}
\dot E^\omega_\infty = \frac{1}{4\pi\omega^2}\int d\Omega\,
|A_4^\omega|^2\;,
\end{equation}
\begin{equation}
e_2^\omega = (\dot E^\omega_\infty)^{-1} \times
\frac{1}{2\pi\omega^2}\int d\Omega\,
|A_4^\omega|^2({\mbox{Re}}\,b_2^\omega)\;.
\end{equation}
In other words, an ${\cal O}(1/r^3)$ correction to $\psi_4$ produces
an ${\cal O}(1/r^4)$ correction to $\dot E$.  We next must understand
how to compute the coefficient of this correction.  We do so using
black hole perturbation theory.

\section{Subleading behavior via perturbation theory}
\label{sec:perturb}

Perturbation theory is a powerful tool for calculating $\psi_4$ and
then determining fluxes such as $\dot E$.  In this section, we use
black hole perturbation theory to confirm the general results of the
preceding section, and to explicitly compute the magnitude of the
subleading contributions to $\psi_4$ and $\dot E$.

Throughout this section we will assume a frequency-domain
decomposition for $\psi_4$.  This assumption means that solutions for
$\psi_4$ separate {\cite{teuk73}}:
\begin{equation}
\psi_4 = \frac{1}{(r - ia\cos\theta)^4}\sum_\omega R_{lm\omega}(r)
S_{lm}(\theta;a\omega) e^{im\phi} e^{-i\omega t}\;.
\label{eq:psi4_decomp}
\end{equation}
The function $S_{lm}(\theta;a\omega) \equiv S(\theta)$ is a
spin-weighted spheroidal harmonic, and is discussed extensively in
Appendix A of Ref.\ {\cite{h00}}.  It satisfies the eigenvalue
relation
\begin{eqnarray}
\frac{1}{\sin\theta} \frac{d}{d\theta} \left(\sin\theta
\frac{dS}{d\theta}\right) + \biggl[ (a\omega)^2\cos^2\theta +
4a\omega\cos\theta -
\nonumber\\
\left(\frac{m^2 - 4m\cos\theta +
4}{\sin^2\theta}\right) + {\cal E}\biggr]S = 0\;.
\label{eq:spheroid}
\end{eqnarray}
In the $a = 0$ limit, ${\cal E} \to l(l+1)$, where $l$ is the usual
spherical harmonic index.

The function $R_{lm\omega}(r) \equiv R(r)$ is governed by
{\cite{teuk73}}
\begin{equation}
\Delta^2\frac{d}{dr}\left(\frac{1}{\Delta}\frac{dR}{dr}\right)
- V(r) R(r) = -{\cal T}(r)\;,
\label{eq:teuk}
\end{equation}
often called the Teukolsky equation.  Here and in what follows $\Delta
= r^2 - 2Mr + a^2$.  Detailed discussion of the source ${\cal T}(r)$
is given in Refs.\ {\cite{teuk73,h00}}.  For our purpose, it suffices
to note that an effective way to solve Eq.\ (\ref{eq:teuk}) is to
first find a homogeneous solution, setting the source ${\cal T}(r) =
0$.  From these solutions, it is fairly simple to build a Green's
function which we integrate over the source to find the particular
solution for our problem.

We show the potential $V(r)$ in the Appendix.  It depends on the
eigenvalue ${\cal E}$ via the parameter $\lambda = {\cal E} -
2am\omega + a^2\omega^2 - 2$.  An important property of $V(r)$ is that
it is long-ranged: as $r \to \infty$, $V(r) \to r^2$.  This makes
computing $R(r)$ for large $r$ difficult.  An excellent way to
circumvent this difficulty is to first solve the Sasaki-Nakamura
equation {\cite{sn82}},
\begin{equation}
\frac{d^2X}{dr_*^2} - F(r)\frac{dX}{dr_*} - U(r) X = 0\;,
\label{eq:sneqn}
\end{equation}
where $r_*(r)$ is the ``tortoise coordinate,''
\begin{equation}
r_* = r + \frac{2Mr_+}{r_+ - r_-}\ln\left(\frac{r - r_+}{2M}\right) -
\frac{2Mr_-}{r_+ - r_-}\ln\left(\frac{r - r_-}{2M}\right)\;.
\label{eq:rstar}
\end{equation}
The potentials $F(r)$ and $U(r)$ are also shown in the Appendix.
Their key property is that, unlike the Teukolsky equation's $V(r)$,
they are short ranged: As $r \to \infty$, $F \to
(\mbox{constant})/r^2$ and $U \to -\omega^2 + (\mbox{constant})/r^2$.
The solutions $X(r)$ thus approach plane waves in the asympotically
flat region, $X(r \to \infty) \to e^{\pm i\omega r_*}$.  Teukolsky
equation solutions can be then be built from Sasaki-Nakamura equation
solutions by the transformation
\begin{equation}
R = \frac{1}{\eta}\left[\left(\alpha + \frac{\partial_r\beta}{\Delta}
\right) \chi - \frac{\beta}{\Delta}\frac{d\chi}{dr}\right]\;,
\label{eq:transform}
\end{equation}
where $\chi = X\Delta/\sqrt{r^2 + a^2}$.  The functions $\alpha(r)$,
$\beta(r)$, and $\eta(r)$ are listed in the appendix.

A more accurate asymptotic form of $X(r)$ is
\begin{eqnarray}
X(r) = A^{\rm out}P^{\rm out}(r)e^{i\omega r_*} +
A^{\rm in}P^{\rm in}(r)e^{-i\omega r_*}\;,
\label{eq:snsoln}
\end{eqnarray}
where
\begin{eqnarray}
P^{\rm in/out}(r) = 1 + \frac{p_1^{\rm in/out}}{\omega r} +
\frac{p_2^{\rm in/out}}{(\omega r)^2} + \frac{p_3^{\rm
in/out}}{(\omega r)^3}\;.
\end{eqnarray}
The coefficients appearing in this expansion are given by
\begin{eqnarray}
p_1^{\rm in} &=& -\frac{i}{2}\left(\lambda + 2 + 2am\omega\right)\;,
\label{eq:p1in}
\\
p_2^{\rm in} &=& -\frac{1}{8}\left[(\lambda + 2)^2 - (\lambda + 2)(2 -
4 a m \omega)\right.
\nonumber\\
& &\left. - 4[am\omega + 3 i M \omega - am\omega(am\omega + 2 i
M\omega)]\right]\;,
\nonumber\\
\label{eq:p2in}
\\
p_3^{\rm in} &=& -\frac{i}{6}\left[4am\omega + p_2^{\rm in}(\lambda -
4 + 2 am\omega + 8 i M \omega)\right.
\nonumber\\
& &\left.
 + 12(M\omega)^2 - 2p_1^{\rm in}\lambda M\omega -(a\omega)^2(\lambda -
 3 + m^2\right.
\nonumber\\
& &\left. + 2 a m \omega)\right]\;;
\label{eq:p3in}
\end{eqnarray}
and
\begin{eqnarray}
p_1^{\rm out} &=& {\bar p_1}^{\rm in} + \frac{\omega c_1}{c_0}\;,
\label{eq:p1out}
\\
p_2^{\rm out} &=& {\bar p_2}^{\rm in} + \frac{1}{c_0}\left[\omega^2
c_2 - \omega c_1\left(p_1^{\rm in} + \frac{i}{2}\right)\right]\;,
\label{eq:p2out}
\\
p_3^{\rm out} &=& {\bar p_3}^{\rm in} + \frac{1}{c_0}\biggl[\omega^3
c_3 - \omega^2 c_2\left(p_1^{\rm in}+ i\right)
\nonumber\\
& &
+ \omega c_1 \left[{\bar p_2}^{\rm in} + \frac{ip_1^{\rm in}}{2} -
\frac{1}{2} + 2 i M\omega (a\omega m - 1)\right]\biggr]\;.
\nonumber\\
\label{eq:p3out}
\end{eqnarray}
The coefficients $c_0$, $c_1$, $c_2$, and $c_3$ appear in the
definition of the function $\eta(r)$, and are given in the Appendix;
overbar denotes complex conjugate.

The condition that radiation be purely outgoing far from the black
hole picks out a solution of the form
\begin{equation}
X = X_\infty P^{\rm out}(r) e^{i\omega r_*}
\end{equation}
as $r \to \infty$.  Performing the transformation
(\ref{eq:transform}), we find that the Teukolsky solution $R(r)$ can
be written
\begin{equation}
R(r) = r^3 Z_\infty Q^{\rm out}(r) e^{i\omega r_*}
\end{equation}
for $r \to \infty$, where $Z_\infty = -4X_\infty\omega^2/c_0$, and
where
\begin{equation}
Q^{\rm out}(r) = 1 + \frac{q_1}{\omega r} + \frac{q_2}{(\omega r)^2}
+ \frac{q_3}{(\omega r)^3} + \ldots\;.
\label{eq:Qdef}
\end{equation}
The coefficients $q_{1,2,3}$ are given in the Appendix.

Now use this solution to examine the flux of energy a finite distance
from the black hole.  Using Eq.\ (\ref{eq:edot_psi}), we find that
\begin{equation}
\dot E(r) = \sum_\omega \dot E_\omega^\infty |Q_\omega^{\rm out}|^2\;,
\end{equation}
where
\begin{eqnarray}
\dot E_\omega^\infty &=& \frac{|Z_\infty|^2}{4\pi\omega^2}\;,
\\
|Q_\omega^{\rm out}|^2 &=& 1 + \frac{6a\omega(a\omega - m) -
  \lambda}{2\omega^2 r^2} + \frac{M(\lambda - 1)}{\omega^2 r^3}\;.
\end{eqnarray}
Notice that the leading correction to the energy flux appears at
${\cal O}(1/r^2)$, in agreement with Eq.\ (\ref{eq:edot_correction}).
Comparing with Eq.\ (\ref{eq:edot_subleading}), we find that the
coefficient $e_2$ which labels the $1/r^2$ falloff is
\begin{equation}
e_2 = \frac{6a\omega(a\omega - m) - \lambda}{2\omega^2}\;.
\label{eq:e2_Kerr}
\end{equation}
Recall that $\lambda$ is related to the spheroidal harmonic eigenvalue
${\cal E}$; cf.\ Eq.\ (\ref{eq:spheroid}) and following discussion.
For Schwarzschild, this correction is particularly simple:
\begin{equation}
e_2(a=0) = -\frac{l(l+1) - 2}{2\omega^2}\;,
\label{eq:e2_Schw}
\end{equation}
where $l$ is the spherical harmonic index associated with the mode
under consideration.

\section{Discussion}
\label{sec:discuss}

In this analysis, we have demonstrated that whenever one extracts
radiation and radiative fluxes at a finite large radius, the
subleading correction to these quantities is at an order ${\cal
O}(1/r^2)$ beyond the leading asymptotic behavior.  Hence, he
correction to the curvature scalar $\psi_4$ is at ${\cal O}(1/r^3)$,
and to the energy flux is at ${\cal O}(1/r^4)$.

Using black hole perturbation theory, we have shown it is not
difficult to calculate the coefficient of the subleading falloff, at
least for a plane wave.  The results we have found are consistent with
the results shown in Table VI of Ref.\ {\cite{skh07}}.  In that paper,
a time-domain code was used to examine radiation from circular orbits.
The time-domain code does not separate the angular behavior, and so
many values of $l$ are included in the analysis simultaneously.  The
radiation tends to be dominated by $l = m$, with important but
decreasing contributions from $l = m+1$, $l = m+2$, etc.  Our
expectation for the Schwarzschild radiation is thus likely to be close
to the prediction from Eq.\ (\ref{eq:e2_Schw}) for $l = m$, skewed
somewhat by contributions from $l = m + 1$.

Let us test that prediction.  Consider first the results for $m = 2$.
If we assume that the waves presented in Ref.\ {\cite{skh07}} for this
case are dominated by radiation in the $l = 2$ and $l = 3$ modes, then
we expect $e_2$ to be between
\begin{eqnarray}
e_2(a = 0, l = m = 2) &=& -2\omega^{-2}\;,\qquad{\rm and}
\\
e_2(a = 0, l = m + 1 = 3) &=&  -5\omega^{-2}\;.
\end{eqnarray}
Table VI of Ref.\ {\cite{skh07}} shows
\begin{equation}
e_2(a = 0, m = 2) = -2.59\omega^{-2}\;,
\end{equation}
in reasonably good agreement with the intuition provided by our
plane-wave expansion.  Table VI also provides Schwarzschild data for
$m = 3$; if those data are dominated by $l = 3$ and $l = 4$, we expect
$e_2$ to be between
\begin{eqnarray}
e_2(a = 0, l = m = 3) &=& -5\omega^{-2}\;,\qquad{\rm and}
\\
e_2(a = 0, l = m + 1 = 4) &=&  -9\omega^{-2}\;.
\end{eqnarray}
Table VI of Ref.\ {\cite{skh07}} shows
\begin{equation}
e_2(a = 0, m = 2) = -6.20\omega^{-2}\;,
\end{equation}
again agreeing reasonably well with the plane-wave expansion.  By
computing the eigenvalues of the spheroidal harmonics for non-zero
spin, one can likewise show that the Kerr values in Table VI agree
reasonably well with the expectation of our plane-wave expansion.

Bear in mind that the numerical magnitude of the correction we derived
strictly applies only for plane-wave expansions.  As such, although we
can provide good {\it post facto} justification of the coefficients of
the subleading falloff, it would be difficult to predict those
coefficients in advance.  To do so, we would need to know the
weighting of the different $l$ modes which contribute to the
radiation.  Our only purpose in analyzing the coefficients shown in
Ref.\ {\cite{skh07}} is to show that the results presented there are
consistent with our results here.

For many calculations, it will not be worthwhile to decompose the
angular distribution of the waves, and thus to compute the subleading
falloff in the manner shown here.  It should be emphasized that the
radial behavior of the falloff is {\it independent} of the $l$ modes
which contribute to the waves.  As such, it would not be difficult to
extract the radiation at several radii and simply fit the coefficient.
That is what was done in Refs.\ {\cite{skh07}} and {\cite{betal08}}.
Implementing such a multi-radius fit should make it possible to more
accurately extract the asymptotic radiation computed by numerical
analysis, potentially reducing errors in such calculations by several
percent.

In general numerical spacetimes, it may be more complicated to take
advantage of this result.  The key ingredient to making the falloff
work as we have discussed is to choose a tetrad such that the Weyl
scalar $\Psi_3 = 0$.  As long as one can perform a null rotation to
put the spacetime into such a ``transverse'' tetrad {\cite{bb02,n05}},
one should find find that subleading corrections to the flux of
radiation fall off as $1/r^3$.  It may be challenging to implement
this rotation for the general case, but the improvement in accuracy
could make it worthwhile.

\begin{acknowledgments}
We thank Pranesh Sundararajan and Gaurav Khanna for valuable
discussions during the formulation of this analysis, and to Bernard
Kelly for asking us about practical issues in applying this result in
the general case.  SAH is also very grateful to Sam Dolan, who pointed
out errors in the subleading corrections to the asymptotic form of the
Sasaki-Nakamura corrections published in Ref.\ {\cite{h00}}, as well
as to Eric Poisson for helpful comments in a very early stage of this
analysis.  LMB was supported by a Theodore Dunham, Jr.\ Grant of the
F.\ A.\ R., and by NSF Grant PHY-0757344, NSF Grant DUE-0941327, and a
NASA EPSCoR RID grant.  SAH was supported by NSF Grant PHY-0449884 and
NASA Grant NNX08AL42G.  SAH also gratefully acknowledges the support
of the Adam J.\ Burgasser Chair in Astrophysics in completing this
paper.
\end{acknowledgments}

\appendix

\section{Functions from black hole perturbation theory}

In this Appendix, we present various functions which arise in black
hole perturbation theory that we need for our analysis.  The functions
$\eta(r)$, $\alpha(r)$, and $\beta(r)$ which appear in the
transformation law (\ref{eq:transform}) are given by
\begin{eqnarray}
\eta(r) &=& c_0 + c_1/r + c_2/r^2 + c_3/r^3 + c_4/r^4\;,
\\
\alpha(r) &=& -\frac{iK(r)\beta(r)}{\Delta^2} + 3i\frac{dK}{dr} +
6\frac{\Delta}{r^2} + \lambda\;.
\\
\beta(r) &=& 2\Delta\left[r - M - 2\Delta/r - iK(r)\right]\;,
\end{eqnarray}
These functions in turn depend on the coefficients
\begin{eqnarray}
c_0 &=& -12i\omega M + \lambda(\lambda + 2) - 12a\omega(a\omega - m)\;,
\\
c_1 &=& 8ia\left[3 a\omega - \lambda(a\omega - m)\right]\;,
\\
c_2 &=& -24iaM(a\omega - m) + 12a^2\left[1 - 2(a\omega - m)^2\right]\;,
\nonumber\\
\\
c_3 &=& 24ia^3(a\omega - m) - 24Ma^2\;,
\\
c_4 &=& 12a^4\;,
\end{eqnarray}
and the function
\begin{equation}
K(r) = (r^2 + a^2)\omega - ma\;.
\end{equation}
Recall that $\lambda = {\cal E} - 2am\omega + a^2\omega^2 - 2$, where
${\cal E}$ is the eigenvalue of the spheroidal harmonic.

The potential $V(r)$ appearing in the Teukolsky equation
(\ref{eq:teuk}) is given by
\begin{equation}
V(r) = -\frac{K^2 + 4i(r - M)K}{\Delta} + 8i\omega r + \lambda\;.
\end{equation}
The potentials $F(r)$ and $U(r)$ appearing in the Sasaki-Nakamura
equation (\ref{eq:sneqn}) are
\begin{equation}
F(r) = \frac{d\eta/dr}{\eta}\frac{\Delta}{r^2 + a^2}\;,
\end{equation}
\begin{equation}
U(r) = \frac{\Delta U_1(r)}{(r^2 + a^2)^2} + \frac{\Delta dG/dr}{r^2 +
a^2} - F(r)G(r) + G(r)^2\;,
\end{equation}
where
\begin{eqnarray}
U_1(r) &=& V(r) + \frac{\Delta^2}{\beta}\biggl[\frac{d}{dr}\left(2\alpha
+ \frac{d\beta/dr}{\Delta}\right)
\nonumber\\
& & - \frac{d\eta/dr}{\eta}\left(\alpha
+ \frac{d\beta/dr}{\Delta}\right)\biggr]\;.
\end{eqnarray}

The coefficients $p_{1,2,3}^{\rm in/out}$ defined in Eqs.\
(\ref{eq:p1in}) -- (\ref{eq:p3out}) are found by requiring that the
solution (\ref{eq:snsoln}) satisfy the Sasaki-Nakamura equation in
each order in $1/r$.  After transforming to the Teukolsky equation
solution $R(r)$, the different orders in $1/r$ are labeled by the
coefficients $q_{1,2,3}$ defined in Eq.\ (\ref{eq:Qdef}):
\begin{eqnarray}
q_1 &=& p_1^{\rm out} - i - c_1\omega/c_0\;,
\\
q_2 &=& -\frac{1}{4c_0^2}\left[-4c_1^2\omega^2 +
4c_0\omega\left[(p_1^{\rm out} - i)c_1 + c_2\omega\right]
\right.\nonumber\\
& &\left. + c_0^2\left[2 + 2ip_1^{\rm out} - 4p_2^{\rm out} + \lambda
+ 6am\omega - 12iM\omega\right.\right.\nonumber\\
& &\left.\left. - 6a^2\omega^2\right]\right]\;,
\\
q_3 &=& \frac{1}{4c_0^3}\left[4c_0 c_1\omega^2\left[(p_1^{\rm out} -
i)c_1 + 2 c_2\omega - 4c_1^3\omega^3\right]
\right.\nonumber\\
& & \left. + c_0^2\omega\left[-4\omega\left[(p_1^{\rm out} - i)c_2 +
c_3\omega\right] + c_1\left[2 + 2ip_1^{\rm out}
\right.\right.\right.\nonumber\\
& &\left.\left.\left. - 4p_2^{\rm out} + \lambda + 6am\omega -
12iM\omega - 6a^2\omega^2\right]\right]
\right.\nonumber\\
& &\left. + c_0^3\left[4p_3^{\rm out} +
2\omega\left[M(5 + \lambda) - 5ia^2\omega\right]
\right.\right.\nonumber\\
& &\left.\left. - p_1^{\rm out}\left[\lambda - 2\omega(3a^2\omega -
3am + 4iM)\right]\right]\right]\;.
\end{eqnarray}

\end{document}